\documentstyle[aps,twocolumn]{revtex}
\begin{document}

\title
{How to Test Stringy Dark Energy? }
\author
{P.H. Frampton}
\address
{Department of Physics and Astronomy, University of North Carolina,
Chapel Hill, NC 27599-3255. }

\maketitle

\begin{abstract}

It is suggested, by using a covariant lagrangian formalism to estimate
the equation of state $w = p/\rho$, that stringy dark energy predicts
$w < -1$, 
a negative pressure larger in magnitude than that
for a cosmological constant or
quintessence. This would lead to a later transition from decelerating
to accelerating cosmological expansion; $w = - 4/3$ 
is briefly considered as one illustrative example.

\end{abstract}

\medskip
\bigskip
\medskip

The last few years have witnessed a revolution in our knowledge
of the universe. Paramount among recent observational discoveries
the phenomenon of dark energy, probably the most surprising discovery
in physics or astronomy since parity violation almost a half a century
ago. As in that case, the explanation of dark energy will surely
impact a wide range of disciplines.

It was recently proposed that the
cosmological dark energy which causes the present accelerated
expansion of the universe may arise from a stringy origin\cite{BFM}.
In particular, in a toroidal closed string universe the
correlation between winding and momentum modes
of closed strings leads in the phase transition at the Hagedorn
temperature to a condensate phenomenon. 

For stringy dark energy, the pressure can certainly be negative because for
exponentially falling $\omega(\kappa)$ the group velocity 
$v_g \propto d\omega(\kappa)/d\kappa$ is negative and can dominate the
pressure expression. The detailed calculation of the value of $w(Z)$  
depends on the Hagedorn phase transition
where the closed strings are strongly interacting and hence impracticable to
compute. Still we can hope to test the approach already by some
estimates. The idea of an exponentially decrease of $\omega(\kappa)$
at large transplanckian $\kappa$ providing a candidate for
dark energy was proposed in \cite{MBK}.

We discuss the pressure estimate first in general terms
followed by a more specific calculation. Finally, the resultant red-shift for
the transition from decelerated to accelerated cosmological expansion is briefly dicussed. 

{\it General dispersion relation.}
~~Consider a spatially flat FRW universe with the spacetime
line element,
\begin{equation}
ds^2 = -dt^2 + a(t)^2 (\vec{dx})^2.
\label{ds2}
\end{equation}
Let $\phi$ be a field satisfying a wave equation that,
either exactly or in the adiabatic approximation, has
plane wave solutions,
\begin{equation}
\phi_{\vec k}(\vec{x}, t) = \exp[i(\vec{k} \cdot \vec{x}
                  - \int^t \omega(k/a(t')) dt'] ,
\label{planewave}
\end{equation}
where $k \equiv |\vec{k}|$.
If the wave equation were Lorentz covariant in the flat spacetime
limit, then one would have $\omega(k/a(t')) =
( k^2/a^2 + m^2)^{1/2}$. We will suppose that $\omega$
has this form when the momentum $k/a(t)$ is small with
respect to the Planck mass, $m_*$ . For larger momenta, we
will suppose that $\omega(c)$ is a decreasing function
of $c\equiv k/a(t)$.

The group velocity, $v_g$, of a wave packet,
\begin{equation}
\phi(x,t) = \int d^3k A(k) \phi_{\vec k}(\vec{x}, t)
\label{wavepacket}
\end{equation}
formed from these plane waves over a narrow range of $\vec{k}$'s near
the value $\vec{k}$, is
\begin{equation}
\vec{v}_g(t) \equiv
a(t) d{\vec x}_{\rm peak}(t)/dt =
(\nabla_{\vec{c}}\,\omega(c))_{\vec{c} = \vec{k}/a(t)}.
\label{groupvelocity}
\end{equation}
It follows that the group velocity is negative if $k$ is larger
than $m_*$. This means that the wave packet moves in
the direction opposite to the momentum, $\vec{k}/a(t)$.
Consider a set of such wave packets contained at time t
within a given comoving volume. When one of the wave
packets passes out of the comoving volume it transfers
momentum into the volume. This means that the pressure
exerted by the energy density inside a comoving volume
on the bounding surface of the volume is {\it negative}.
As in kinetic theory, one may calculate the momentum
transfer per unit time and per unit area. This gives
the pressure $P$ exerted by such wave packets as,
\begin{equation}
P(t) = \frac{N}{3} \int d^3k\, n(c(t)) \vec{c(t)}\,\cdot \vec{v}_g (c(t))
\label{pressure}
\end{equation}
Here $n(c(t))$ is the number per unit physical volume,
of wave packets having momentum $c(t)$ at time $t$. This
number density is uniform at any given time.
Similarly, one finds that the energy density $\rho$
of these wave packets is
\begin{equation}
\rho(t) = N \int d^3k\, n(c(t))\, \omega(c(t)).
\label{energydensity}
\end{equation}

The packets with momenta that satisfy $c(t) > m_*$
exert negative pressure, but have positive
energy density. Therefore, they may contribute to the
recent acceleration of the universe. Packets that
satisfy this condition, correspond to modes with
$k > m_* a(t)$. A mode, $\vec{k}$, that satisfies
this condition at the present time, had to have
transplanckian momenta in the very early universe.
Since $\omega(c(t))$ decreases with increasing values
of $c$ larger than $m_*$, the energy $\omega$ of such
a packet was less in the early universe than it is today.
This is consistent with negative pressure, which causes
the energy density to increase as the universe expands.
The pressure and energy density of these packets satisfies
the local conservation law stated in Eq.(\ref{conserve}) below.

A given mode $\vec{k}$ that has momentum
$k/a(t) \gg m_*$ will increase in energy $\omega(k/a(t))$
as the universe expands and will exert negative
pressure until the momentum becomes less than $m_*$. After
that, the energy and pressure of the mode behaves more like
that of an ordinary particle.

The pressure and energy density here refer to the specific
coordinate system in which the preferred geodesics correspond
to world lines of constant $\vec{x}$. Thus, the meaning of
transplanckian momentum is defined with respect to the
set of preferred geodesics. Furthermore,  $P$ and $\rho$
are the pressure and energy density that would be measured
by an observer on one of these geodesics. Thus, they are
scalar quantities. The corresponding energy-momentum tensor
for this set of packets is therefore
\begin{equation}
T_{\mu\nu} = P g_{\mu\nu} + (\rho + P) u_{\mu} u_{\nu}
\label{energymomentumtensor}
\end{equation}
Then Eq. (\ref{conserve}) is equivalent to the local
conservation law, $T^{\mu\nu}{}_{;\nu} = 0$, for the metric
of Eq. (\ref{ds2}), as well as in any spatially curved
FRW universe.

The extension of this discussion to spatially curved FRW
universes is straightforward, with the spatial part of the
mode functions,$\exp(i\vec{k}\cdot \vec{x})$, replaced by
the appropriate harmonic functions that are eigenfunctions
of the spatial Laplacian operator.

In an FRW spacetime having  metric of Eq. (\ref{ds2}),
once the energy function $\omega$ is chosen, one
can calculate its effect on the scale factor $a(t)$ through
the Einstein equations,
\begin{equation}
R_{\mu\nu} - (1/2) g_{\mu\nu} R = 8\pi G T_{\mu\nu}
\label{einstein}
\end{equation}
where now the energy-momentum tensor includes all
contributions. The negative pressure coming from the
modes with transplanckian momenta will tend to accelerate
the expansion of the universe. The degree of acceleration
will depend the density and pressure of other forms of
matter present and on the rate at which $\omega$ decreases
with increasing momentum in the transplanckian range.
One simple example of a possible $\omega$ is
\begin{equation}
\omega(c) = \sqrt{\vec{c}^2 + m^2} \exp(-c/m_*),
\label{exponentialomega}
\end{equation}
where m is the mass of the field $\phi$
and $m_*$ is the Planck mass.
In the next section, we discuss a Lagrangian which gives rise
to a particular dispersion relation $\omega$
that has a form similar to Eq. (\ref{exponentialomega}),
but with $\exp(-c/m_*)$ replaced by $\exp(-c^2/m_*^2)$.
In addition, the Lagrangian we discuss treats $u^{\mu}$
as a dynamical field.

{\it Equation of State for Stringy Dark Energy.}
~~We define a dimensionless $\kappa = k/M_{Planck}$ scaled relative
to the Hagedorn temperature which we designate as $M_{Planck}$.
We further define the comoving momentum
$c = \kappa /a(t)$ where $a(t)$ is the cosmological
scale factor.

The dispersion relation in \cite{BFM} is linear for small $k$, 
$\omega(c)^2 \sim c^2$
as is necessary to reproduce 
scale-invariant large-scale fluctuations as well as the
black-body CMB spectrum. At very high $c$, $\omega(c)$ falls
as a gaussian{\footnote{Note that in \cite{BFM}
an exponential $\omega(c) = ce^{-\gamma c}$
rather than a gaussian form
$\omega(c) = ce^{-\gamma c^2}$
was discussed but the qualitative features
of the putative dark energy candidate
remain unchanged; we find the gaussian form
more convenient to develop
a covariant lagrangian formalism}.
We parametrize these features by the form
\begin{equation}
\omega(c) = c e^{- \gamma c^2}
\label{dispersion}
\end{equation}
where $\gamma \sim O(1)$ is real.

To estimate the equation of state, we find it
most convenient to employ the formalism of
general relativity in a preferred frame as is necessary, for
example, to underwrite the non-gravitational theories
\cite{CG} of Lorentz violation. 

We adopt an action:

\begin{eqnarray}
S & = &  \int d^4x \sqrt{-g} 
\left[ R + {\cal L}_u + \lambda (u_a u^a - 1) + {\cal L}_{\phi}
\right] 
\label{action}
\end{eqnarray}

\noindent with

\begin{equation}
{\cal L}_{\phi}  =  \partial_{\mu}\phi \partial^{\mu} \phi +
\sum_{n=1}^{\infty} \frac{(- \gamma)^n}
{n! M_{Planck}^{2n}} 
D^{2n}\phi D^2 \phi 
\label{Lphi}
\end{equation}
The four terms in $S$ require further 
explication. The first is the Einstein-Hilbert action.
In the second, $u^a$ is a comoving timelike
vector, specifying the preferred frame for FRW cosmology,
necessarily promoted to the status of a field to
allow consistent conservation laws.
For ${\cal L}_u$ we may write, for example, a form
similar to Eq.(\ref{Lphi}):
\begin{equation}
{\cal L}_{u}  =   
\sum_{n=0}^{\infty} \frac{(- \beta)^n}
{n! M_{Planck}^{2n}} 
D^{2n}u_a D^2 u^a 
\label{Lu}
\end{equation}
Note that Eq.(\ref{Lu}) is merely illustrative
and physically reasonable, although our conclusions do
not depend on a specific choice for ${\cal L}_u$.
It is important that we do not
specialize to the Yang-Mills-like term
\cite{JM1,JM2} ${\cal L}_u \propto F^{ab} F_{ab}$ with 
$F_{ab} = 2\nabla_{[a} u_{b]}$
since this would lead to the unacceptable result for
$w = <P>/<\rho>$
found in \cite{LeM}.
In our analysis the more general
Eq.(\ref{Lu})
is crucial in
obtaining a physically-meaningful
equation of state.
In the third term
of Eq.(\ref{action}), $\lambda$ is a lagrange multiplier
for the constraint that $u^a$
be of unit lengthad have squared norm.
The fourth and last term in Eq.(\ref{action})
is a choice for ${\cal L}_{\phi}$
which leads to the 
form of modified dispersion relation
$\omega(c) = c exp (- \gamma c^2)$.
Here $D^2 \phi = - D^a D_a \phi 
= - q^{ac} \nabla_a (q^b_c \nabla _b \phi)$
with $q_{ab}$ the spatial metric orthogonal
to $u^a$: $q_{ab} = - g_{ab} + u_a u_b$.

The contributions to the stress-energy
tensor are from $T_{ab}^{(u)} + T_{ab}^{(\phi)}$.
We find now that
the pressure $<P> = \frac{1}{3} <T_{ii}>$ receives a contribution 
from $T_{ii}^{(\phi)}$ from which we therefore derive the result: 

\begin{eqnarray}
<P> 
& = & 
\frac{N}{3} \int 
\frac{dc c^2}{\omega(c)} n(c)
\left[\omega(c)^2 + 
2\sum \frac{(\gamma c^2)^{n+1}}{n!} 
\right.
\nonumber \\
& & \left. - 2 \sum 
\frac{(n+1)(\gamma c^2)^{n+1}(-1)^{n+1}}{n! }
\right]
\label{press}
\end{eqnarray}
and a similar contribution from $T_{ii}^{(u)}$.
We may re-write the pressure as follows

\begin{eqnarray}
<P> 
& = &
\frac{N}{3} \int_{c_H}^{\infty} c^2 dc 
n(c) \left[ c \omega^{'}(c) + u_a \bar{\omega}^{'a}(u)  \right]
\label{PR}
\end{eqnarray}
Here $\bar{\omega}$ is defined as the differential of
Eqs.(\ref{action}) and (\ref{Lu}).
In Eq.(\ref{PR}), the factor $\frac{1}{3}$ arises from
the spatial dimension and the group velocity
$v_{group} \propto c d\omega/dc$ is occurring naturally.
The notation is $\omega^{'} = d\omega/dc$ and
$\bar{\omega}^{'a} = d\bar{\omega}/du_a$.
Our model has some superficial similarity to, but is quite different from,
k-essence\cite{kessence}.

Now we need the density $<T_{00}>$ which
receives contributions from
$T_{ab}^{(\phi)}$ and $T_{ab}^{(u)}$
in the general case.
One finds by a similar analysis to the above
for pressure that the mean density is

\begin{equation}
<\rho> = N \int_{c_H}^{\infty}
c^2 dc n(c) ( \omega(c) + \bar{\omega}(u))
\label{rho}
\end{equation}

We next confirm that Eq.(\ref{PR}) and Eq.(\ref{rho})
satisfy conservation of energy for general $\omega$
and $\bar{\omega}$, thus confirming that ${\cal L}_u$
as illustrated in Eq.(\ref{Lu}), is not constrained.
We assume $\alpha$ and $\beta$ are of order one and that the
occupation numbers satisfy $n(c) \simeq n(u) \sim O(1)$.
We compute the conservation law as follows:

\begin{equation}
\rho =  \frac{1}{a^3} (\omega + \bar{\omega}) ~~~{\rm and}~~~
P = \frac{1}{3 a^3} (c \omega^{'} + u_a \bar{\omega}^{'a}) 
\end{equation}
satisfy $TdS = dE + PdV = 0$, that is
\begin{equation}
\frac{d}{dt} (\rho a^3) + P \frac{d}{dt} (a^3) = 0
\label{conserve}
\end{equation}
providing that
\begin{equation}
\dot{\omega}(c) = - H c \omega^{'} ~~~{\rm and} 
~~~ \dot{\bar{\omega}} = - H u_a \bar{\omega}^{'a}
\label{consistency}
\end{equation}
Given this consistency condition, an arbitrary modified
dispersion relation $\omega$ and arbitrary
choice of $\bar{\omega}$
satisfy conservation of entropy.

In summary of the equation of state, we have:

\begin{eqnarray}
w & = & 
\frac{ \frac{1}{3}\int_{c_H}^{\infty} dc c^2 n(c) (c\omega^{'}(c) 
 + u_a\bar{\omega}^{'a}(u))}
{\int_{c_H}^{\infty} dc c^2 n(c) (\omega(c) + \bar{\omega}(u))}
\label{w}
\end{eqnarray}

\noindent With the illustrative choice given earlier
one has $c \omega{'} = (1-2\gamma c^2)\omega$
and $u_a\bar{\omega}^{'a} = (1 - 2\beta u^2)\bar{\omega}$
so for this case if $\beta < a^2/2$ there is a positive contribution
to pressure from the $u$ term and, as explained above,
a negative pressure expected to dominate physically
from the $c$ term.

In the absence of an explicit non-adiabatic computational framework 
let us assume that $w < -1$ but is not unacceptably-negative.
The important and attractive point is that the order-of-magnitude estimate
for the dark energy density giving $\Lambda \simeq 10^{-120}$
in Planckian units\cite{BFM} is preserved even when 
we arrive at a sensible equation of state which, from the form
of Eq.(\ref{w}), still satisfies $w < -1$. A particular solution
which asymptotes to $w = -1$ is discussed in \cite{BM}.

Let us now consider such a case with $w < -1$. As an interesting example 
consider the case where the stringy dark energy density $\rho_{DE}$ 
increases with $(1 + Z)^{-1}$
corresponding to $w = -\frac{4}{3}$.
We note that a different theory predicting $w < -1$, 
based on a quantized very-light free scalar field, was considered in \cite{parker}
and that the observational status of $w < -1$ was discussed in \cite{caldwell}.

{\it Transition from Deceleration to Acceleration.}
~~It is suggested by the study of Supernova 1997ff \cite{riess}
that the present accelerated expansion of the Universe with
deceleration parameter $q_0 < 0$ commenced at a finite transition redshift
$Z_{tr}$ when $q(Z_{tr}) = 0$.
Another key red-shift is $Z_{eq}$ when the dark energy begins to
dominate the total energy density;
assuming flatness $\Omega_{\Lambda}(Z) + \Omega_{M}(Z) = 1$,
there was equality $\Omega_{\Lambda}(Z_{eq}) = \Omega_{M}(Z_{eq}) = 1/2$.

We adopt the most favored values for the present matter and dark
energy content in terms of the critical density
$\rho_{cr} = (3 H_0^2 / 8 \pi G)$ 
as $(\Omega_{DE})_0/(\Omega_M)_0 = 2$.
These are suggested by the common intersection of three
sources: the supernovae type IA data\cite{perlmutter,riess2},
the observations of the acoustic peaks in the Cosmic Microwave Background
from BOOMERANG\cite{boomerang} and MAXIMA\cite{maxima} reviewed in \cite{CMBreview}
and the estimates of dark matter
in large scale studies. For an overview, see\cite{Costri}.
The values of $Z_{tr}$ and $Z_{eq}$ satisfy
\begin{equation}
1 + Z_{tr} = \left( -(1 + 3w) \frac{(\Omega_{DE})_0}{(\Omega_{M})_0} \right)^{-1/3w}
\end{equation}
\begin{equation}
1 + Z_{eq} = \left( \frac{(\Omega_{DE})_0}{(\Omega_{M})_0} \right)^{-1/3w}
\end{equation}
where $w = <p>/<\rho>$ for the dark energy.
Putting in $w = -4/3$ as in our example predicts $Z_{tr} = 0.57$ and $Z_{eq} = 0.19$.
For comparison, these values for a cosmological constant are $Z_{tr} = 0.59$ and
$Z_{eq} = 0.26$ while for quintessence models with $w > -1$ the
corresponding $Z_{tr}, Z_{eq}$ values are different again.

From data on radio galaxies\cite{Daly} there is
some preliminary indication that $w < -1$ might be favored.
Also, a global fit to several cosmological data sets gives
\cite{HM} a range $-2.68 < w < -0.78$ which implies that
$w < -1$ is still a viable possibility,
This inequality may be the first support for a stringy 
origin to the puzzling phenomenon
of dark energy.

\bigskip
\bigskip
\bigskip

I wish to thank L. Mersini and L. Parker for discussions.
This work was supported in part by the
Office of High Energy, US Department
of Energy under Grant No. DE-FG02-97ER41036.

\end{document}